\documentclass[aps,prb,twocolumn,preprintnumbers,amsmath,amssymb]{revtex4}
\usepackage{dcolumn}% Align table columns on decimal point
\usepackage{bm}% bold math
\usepackage{amsmath}
\usepackage{feynmp}
\usepackage{graphicx}% Include figure files

\begin{document}
\title{Indications for Topological Physics in High-$T_c$ Materials}
\author{Vincent E. Sacksteder IV$^{1}$}
\affiliation{W155 Wilson Building, Royal Holloway University of London, Egham Hill, Egham, TW20 0EX, United Kingdom }
\affiliation{Department of Physics and Astronomy, Rutgers University, New Jersey 08854, USA}
\email{vincent@sacksteder.com}

\date{\today}

\begin{abstract}
Research on high-$T_c$ superconductors has generally not focused on analysis of the topological structure of electronic bands in these materials.  In this article we collate and discuss several well-known experimental  observables that  signal that certain signatures of topology may be present.  The topological signatures suggested by experiment include dimensional reduction, electronic transport determined by loop statistics, a winding number, % a geometrical quantity [2-D sheet density] controlling characteristic temperatures,   
Weyl fermions, and others.   We suggest that topology,  in a wider sense than band topology, may be key to the mechanism of high-$T_c$ superconductivity.  
\end{abstract}
 
\maketitle

Topology is the study of those properties of geometrical objects which do not change when an object is deformed, but do change when it is torn or glued together. For 1-D curves topology studies their knots but not their length or shape.  For a 2-D surface topology may study the edges and corners and punctures through the surface, while for 3-D volumes the volume's surface, edges,  corners and again punctures may be of interest.  

Topology has been  used in modern condensed matter physics to understand topological insulators, Weyl fermions, and Dirac cones.  Here topology is often meant and used in a  way that is much more restricted than its general mathematical meaning.  We will call this restricted version band topology.   Band topology is rooted in a perspective that starts from momentum space, and is used to analyze the band structure that results when a material obeys  lattice translational symmetry.  Depending on the symmetries obeyed by a material, its band structure may be characterized by a topological invariant, an integer which is protected by a gap between the bands in the bulk of the material. The topological invariant cannot change unless the bulk band gap closes and the neighboring bands meet each other.  As long the gap remains open and the topological invariant retains a non-zero value, it is guaranteed that conducting states will remain on the surface [in real space] of the material and that the surface will remain conducting.  This style of topological analysis based on band structure, i.e. band topology, often assumes that scattering, disorder, electronic correlations, and electronic interactions are in some sense weak. \cite{RevModPhys.82.3045,RevModPhys.83.1057,RevModPhys.88.035005,RevModPhys.88.021004,RevModPhys.90.015001}

Topology has also been used  to give an account of  electronic conduction in specifically 2-D geometries, where the integer Quantum Hall effect (IQHE) and fractional Quantum Hall effect (FQHE) are found.  The same combination of topology and momentum space, i.e. band topology, that we mentioned above can be applied to the IQHE. \cite{PhysRevLett.49.405}  However in these systems a different style of topological analysis, based in real space, has also been successful.  Success has been obtained both when interactions are weak and when they are strong. The Laughlin wave-function pioneered this success when interactions are strong.  Vortex features of the wave-function in real space, combined with the specifically two dimensional geometry, have been key to this real-space topological analysis. \cite{cage2012quantum,RevModPhys.71.S298,pruisken2010topological} 

The physics community has in general shown little temptation to  look for connections between band topology and the world of high $T_c$ superconductivity and bad metals. In these materials any connection to Fermi liquids and band structure is suspect, and therefore looking at topological properties of the band structure does not seem promising.  
%There are no special indications that these materials belong to a special symmetry class suitable for realizing a topological insulator. 
Instead the focus of research has been on the nature of the pairs which mediate high $T_c$ superconductivity,  the pairing mechanism which binds those pairs, various orderings such as charge density waves, and the nature of the strange metal state and pseudogap seen at temperatures above $T_c$.  Neither the topological features of the quantum wave-function, nor the machinery of band gaps, topological invariants, and Dirac dispersions,  seem likely to cast much light on these  questions.  \cite{RevModPhys.83.1589,keimer2015quantum,stewart2017unconventional}

In the particular case of the iron pnictide superconductors, there has been some interest in the presence of Dirac cones and topological phases, but the emphasis  has been on exploiting the combination of topology and superconductivity for new technologies. \cite{PhysRevLett.105.037203,PhysRevLett.104.137001,bhoi2011quantum,PhysRevB.84.100508,PhysRevB.88.165402,pallecchi2013role,PhysRevB.90.024517,PhysRevB.93.104502,PhysRevB.93.104513,PhysRevB.96.220509,PhysRevLett.119.096401,PhysRevX.8.011014,zhang2019multiple,hao2019topological}

Nonetheless there are several signs  that  topology, in a broader sense than the band topology frequently used in condensed matter physics, may be key to understanding high-$T_c$  superconductivity and bad metals.  The goal of this article is to collect those signs known to the author into one place,  with some brief but not terribly technical discussion.  The hope is not to prove  beyond every doubt the relevance of topology to high $T_c$, but instead to gather the evidence and provoke new thinking and inquiry.

Section \ref{FermiArcs} of this article discusses experimental reports of Fermi arcs in the cuprates, which if taken very naively indicate the presence of Weyl fermions. Next section \ref{PseudogapHierarchy} describes a pattern in the doping vs. temperature phase diagram of two cuprates, LSCO and YBCO.  In both materials several  characteristic temperatures, including the pseudogap temperature, decrease linearly with doping and converge at a common meeting point.  In LSCO there are four such temperatures, while in YBCO there are at least three such temperatures. This pattern  suggests that a topological winding number is at work.  Section \ref{DimensionalReduction} presents                                                                                                                                                                                                                                                                                                                                                                                                                                                                                                                                                                                                                                                                                                                                                                                                                                                                                                                                                                                                                                                                                                                                                                                                                                                                                                                                                                                                                                                                                                                                                                                                                                                                                                                                                                                                                                                                                                                                                                                                                                                                                                                                                                                                                                                                                                                                                                                                                                                                                                                                                                                                                                                                                                                                                                                                                                                                         two signs in the cuprates of dimensionality being reduced below the nominal 3-D character of the cuprate materials; reduced dimensionality is a hallmark of topological states.  Section \ref{TISimilarities} discusses three qualitative similarities in electronic transport between topological insulators and high-$T_c$ materials.  Sections \ref{DimensionalReduction}, \ref{TISimilarities}, and  \ref{LoopKeyRole} analyze the linear resistance (linear in temperature and in magnetic field) seen in high $T_c$ materials, and the connection to loops, which are topological objects. Section \ref{LinearRelations}  widens this analysis to several other linear relations seen in high-$T_c$ materials, including temperature vs. the 2-D sheet density of holes, and temperature vs. superfluid phase stiffness.   Section \ref{AtomicUnits} discusses experimental results on numerical values of the slopes of these linear relations, and also the ratio of a material's $T_c$ to its 2-D density of holes at optimal doping.  In atomic units these experimental values are all dimensionless and of order $O(1)$, suggesting that atomic units are somehow preferred by the physical mechanism of high-$T_c$.  We discuss the possibility that this preference may reflect topological physics connected with  electron-hole loops.   Section \ref{Meissner} gives a brief comment on the Meissner effect.  Lastly Section \ref{Postscript} is a postscript listing several strategies one might use to do topology in strongly correlated materials.

\section{Fermi Arcs\label{FermiArcs}}
Many experiments on cuprates have reported that the Fermi surface is in fact four Fermi arcs:  four curves which each terminate at two end points, with the end point location changing with temperature and doping. \cite{norman1998destruction,vishik2018photoemission,yazdani2016spectroscopic,kordyuk2015pseudogap}  Within a weakly interacting approach it is impossible to account for Fermi arcs that are truly incomplete, at least in the bulk in the material.  The difficulty is that the Fermi surface defines a boundary between the part of the Brillouin zone where all states are occupied and the part of the Brillouin zone where all states are unoccupied.  If the Fermi surface is incomplete then it cannot be interpreted as the boundary between occupied and unoccupied sectors.  Therefore it is widely understood that Fermi arcs, if they do exist in the cuprates, are a manifestation of strongly correlated physics.  The details of how such physics would result in Fermi arcs are not well understood or agreed on. 

There is, however, a well-understood way to explain Fermi arcs  within the theory of weakly interacting materials and band structure. This explanation is a topological explanation, and it puts the Fermi arcs on the material's surface rather than in its bulk.  The theory of how such Fermi arcs arise has a solid foundation, and has been verified by experimental observation of Fermi arcs in several topological materials.  \cite{RevModPhys.90.015001}  Since there is a clear theory and experimental realization of Fermi arcs in topological materials, and the theories of strongly interacting Fermi arcs are on a weaker footing, some attention should be be given to the possibility that the Fermi arcs seen in the cuprates are caused by topology.  This possibility apparently has not been discussed, so here we breach the question, putting it baldly. \footnote{We did many detailed and lengthy searches of citations of the original article on Fermi arcs in cuprates \cite{norman1998destruction}, and found no discussion of the possibility that these might be associated with Weyl fermions. } A hypothesis can not be proven wrong until it has been considered.

The cuprates' Fermi arcs are seen at a range of temperatures above the superconducting temperature $T_c$ and below another temperature $T^*$ which decreases with doping.  At temperatures above $T=T^*$ there is a single Fermi surface, a circle around the $\left[\pi,\pi \right]$ point in the Brillioun zone.  This point   is equivalent also to three other points $\left[\pm \pi, \pm \pi \right]$.   At $T=T^*$ the circle  separates into four arcs, with pieces missing near the boundaries of the Brillouin zone.  The arc length decreases with temperature until at $T=T_c$ there remain only four Fermi points.  The arc length also depends on doping.  

There are several difficulties with analysis of the Fermi arcs: (1) They are observed using surface not bulk spectroscopy: ARPES or STM. These experimental tools do not see into the bulk.  (2)  It is very hard to get clean stable surfaces in  La$_{2-x}$Sr$_{x}$CuO$_4$ (LSCO) or  YBa$_2$Cu$_3$O$_{6+\delta}$ (YBCO), so almost all experimental observations of Fermi arcs have been in Bi$_2$Sr$_2$Ca$_{n-1}$Cu$_n$O$_{2n+4+x}$ (BSCCO).  (3) ARPES and STM report results that do not entirely agree with each other.  (4) Doping is hard to control so it is hard to  measure the doping dependence of the Fermi arcs.  (5) Most of the spectral weight seen in ARPES is spread smoothly in energy, as is typical in strongly correlated materials.  The Fermi surface signal strength, i.e. the height in the small peak interpreted as a Fermi surface, is a small fraction of the smooth background.  Therefore a lot depends on how one fits the signal.  (6) Some  claim that the disappearance of portions of the Fermi surface is not real - it is caused by broadening [associated with temperature and interactions] which washes out the peak in the spectral density; if the broadening did not occur then the Fermi surface would remain intact.  Others claim  that between the two endpoints of the Fermi arc there is a second connecting arc which for various reasons is less visible to ARPES and STM, so that there are four complete Fermi surfaces, and no real Fermi arcs.  \cite{vishik2018photoemission,yazdani2016spectroscopic,kordyuk2015pseudogap}

We leave aside these concerns and take a naive approach to the overall story of Fermi arcs, taking at face value the claim that the Fermi surface in BSCCO really is incomplete.    

The only way to explain Fermi arcs, within a weakly interacting approach, is  if they reside not in the material's bulk but instead on its surface.    Certain topological materials  host in their bulk Weyl fermions, i.e. fermions with a well-defined chirality [either right-handed or left-handed] and obeying a Dirac cone linear energy dispersion.  Weyl fermions always come in pairs, with the pair separation [within momentum space] determined by  the strength of time-reversal symmetry breaking via a magnetization or magnetic field.  \cite{PhysRevB.94.085128} [Inversion symmetry breaking is also an alternative to  time reversal symmetry breaking.] Any number of Weyl fermion pairs can be realized in a material.   On the surface of a Weyl material Fermi arcs connect each pair of Weyl fermions.  \cite{RevModPhys.90.015001}

In BSCCO there are four Fermi arcs, each with two end points.  
If the Fermi arcs seen in BSCCO are the surface manifestation of bulk Weyl physics, then BSCCO must host eight Weyl cones in its bulk.

There is no conceptual difficulty with producing a Hamiltonian which has the same four Fermi arcs and eight Weyl cones seen in BSCCO.  A simple model with the copper oxide layer's $C_4$ symmetry and  four Fermi arcs arranged around $\left[\pm \pi, \pm \pi \right]$ has already been reported - see Fig. 3-d1 in Ref.  \onlinecite{PhysRevB.96.235424}. Exact reproduction of the experimental data on cuprate Fermi arcs and their dependence on doping and temperature is simply a matter of choosing a convenient Hamiltonian. The Hamiltonian's parameters must be tuned to depend on temperature and doping,  in order to  reproduce the fragmentation of the Fermi surface into Fermi arcs at the pseudogap temperature, and also to reproduce the reduction of the arcs to nodes at $T_c$.

In summary, if the experimental evidence for Fermi arcs is taken naively, then it implies that BSSCO at least, and possibly all the cuprates, are Weyl materials in the bulk, as long as the temperature lies above $T_c$ and below the pseudogap temperature $T^*$.  In this naive interpretation,  ARPES and STM measurements on BSCCO are reporting surface not bulk Fermi arcs. 

Section \ref{Postscript} has some brief comments about how Weyl fermions might arise at long distance scales as the result of renormalization group flow from a non-Weyl Hamiltonian.

\section{The Hierarchy of Pseudogap Lines \label{PseudogapHierarchy}}
This author recently performed a comprehensive survey of all experimental reports of characteristic temperatures   well above the superconducting transition $T_c$ in strontium doped lanthanum cuprate (LSCO) and oxygen doped YBCO.  \cite{sacksteder2020quantized} We report  only  characteristic temperatures at which  clearly identifiable signals occur, for example opening of a pseudogap, peaks or kinks in the temperature dependence of various observables, or extinction of a diffraction peak at a particular temperature.  These characteristic temperatures have been found using a wide variety of measurement techniques over a period of about three decades.   The reported temperatures have been realized experimentally, rather than being derived by extrapolation from lower temperatures.  We used only characteristic temperatures that were reported numerically in the original reports, rather than extracting temperatures from data ourselves.  [There was one exception to this rule.] We omitted the experimental literature that focuses on ionic behavior, but were very thorough about including all other experimental data sets.  The experimental corpus on characteristic temperatures from 1990 until 2018 contains twenty separate data sets from thirteen experimental groups on LSCO, and twenty-six distinct data sets on YBCO.

When the experimental corpus for LSCO is plotted all on one graph, with temperature as the $y$ axis and hole doping as the $x$ axis, a remarkable pattern emerges.  The reported characteristic temperatures lie on a sequence of distinct well-separated straight lines which radiate from a common intersection near the high-doping end of the superconducting dome. The highest line is associated with a peak in the magnetic susceptibility and transport signatures, the second line has symmetry breaking from tetragonal to orthorhombic crystal structure, the third line is the pseudogap transition seen in ARPES, and the fourth line is seen in transport signatures and NMR.   There is some evidence suggesting lower lines as well. We call this family of lines the pseudogap family.

This family of pseudogap  lines in LSCO obeys a quantization rule: starting from the highest characteristic temperature, the second temperature's slope and value are one half those of the  highest temperature, the third line's slope and value are one third, and the fourth line's slope and value are one fourth. 

A similar pattern is visible in YBCO.  Here three well-separated lines are visible, again with a common intersection at high doping. The highest line is the pseudogap transition accompanied by symmetry breaking, the next shows nematicity in transport and also time reversal symmetry breaking, and the  next is seen in transport signals. The slopes of the three YBCO lines have the  the same fractional relation to each other as is seen in the second, third, and fourth lines in LSCO.  This suggests the existence of another higher line at twice the temperature of the highest observed line.  There is some data suggestive of the higher line, but it is sparse. Possibly this is because YBCO is rarely measured above $300\,K$,  unlike LSCO which has been measured at temperatures as high as $T = 700\,K$.    In any case the family of  pseudogap lines  observed in YBCO show the same common intersection and quantized slopes that are seen in LSCO's pseudogap family.

A pattern of lines with quantized $1/n$ slope and common intersection is known in another context: the energies of Landau levels of a  fermion moving in a plane, with a Berry phase like that caused by a Dirac point.  \cite{PhysRevLett.82.2147,PhysRevB.82.085429} The analogy between the pseudogap lines and the  Landau levels would require that hole doping map to the Dirac fermion's Fermi energy, and that temperature map to the Dirac fermion's magnetic field $B$.  
 While this mapping is unmotivated and obscure, the observed pattern of quantized lines suggests  a $U(1)$ phase and a winding number associated with that phase, i.e. the winding number is the number of times that the phase goes through $2 \pi$.  For Landau levels the $U(1)$ phase is of course tied to magnetic flux.  Winding numbers are one of the most elementary realizations of topology seen in physics. If the hierarchy of pseudogap lines seen in LSCO and YBCO really is caused by winding number physics, this would explain why the observed  family of pseudogap temperatures  is so robust against interactions. 
%Results on the Landau levels:
%3-D sigma.p : E = \pm v_F \sqrt(p_z^2 + 2 B n), abrikosov prb 58.2788
%3-D p^2/2m: (n+1/2) B + p_z^2/2m
%2-D sigma.p + p^2/2m + g B \sigma_z  : E = B (n \pm (1/2) \sqrt{(1-g)^2 + 8 n v_F^2 / B}, where g is the strength of the Zeeman term. prb 82 085429.  see also hanaguri E = v_f \sqrt{2 n B}, prb 82 081305(R)
%2-D p^2/2m: (n+1/2) B, and if there is a Berry phase(for instance caused by a Dirac point i.e. a protected band crossing) then the 1/2 should be removed.

In addition to the possible winding number discussed above, it is worth dwelling a little more on the fact that the slopes of the pseudogap lines seen in LSCO and YBCO follow a $1/n$ pattern of $1/1, \, 1/2, 1/3, \, 1/4$.  2-D Landau levels reproduce this pattern only if the electronic motion is affected by either a Zeeman term or a  Berry phase like that seen when there is a topologically protected Dirac point.    \cite{PhysRevLett.82.2147}  Otherwise 2-D Landau levels follow a $1/(n+1/2)$ pattern: $(1/2)^{-1}, \, (3/2)^{-1}, \, (5/2)^{-1}, \, (7/2)^{-1}$.  The experimental data do not support this pattern and support the $1/n$ pattern instead. The existence of a $1/n$ not $1/(n+1/2)$ pattern in the pseudogap lines may be a hint of topological physics in the cuprates.

\section{Dimensional Reduction \label{DimensionalReduction} }
One hallmark of topology is reduced dimensionality.  In a 2-D material topological states appear at the edges and corners, while in a 3-D material topological states appear at the surface, edges, and corners.  When the key physics happens at a  dimensionality that is smaller than the apparent dimensionality of a system, this can indicate that topology is at work.

\textbf{The Copper Oxide Plane.} In this respect it is interesting that the essential physics of the cuprates lies  in the copper oxide CuO$_2$ plane, even though the actual material is 3-D.    Some variation in $T_c$ occurs as the number of CuO$_2$ planes is varied, but the variation in $T_c$ is relatively weak, suggesting that it is a second order effect.   In LSCO  all the physics required for its high-$T_c$ superconductivity  lies in the plane, since  monolayer LSCO reaches the same $T_c$ as bulk LSCO.   \cite{bovzovic2016dependence} The focus on 2-D physics is also reminiscent of the IQHE and FQHE systems, which also work on topological grounds.

\textbf{Linear Resistance in Bad Metals May be a Sign of Quasi-1-D Conducting Channels.}
At temperatures above $T_c$ the cuprate and pnictide superconductors have a resistance that increases linearly with temperature, and this linear growth continues to large temperatures without saturating.  Linear-in-temperature resistance has become known as a hallmark of strong correlations, including in materials that do not superconduct. Materials with this signature are called bad metals. \cite{hussey2004universality}

In recent years experiments have found that several materials with linear-in-temperature resistance also exhibit linear-in-magnetic field resistance. This was verified in the high-$T_c$ cuprate superconductor La$_{2-x}$Sr$_{x}$CuO$_4$  (LSCO), the high-$T_c$ pnictide superconductor BaFe$_2$(As$_{1-x}$P$_x$)$_{2}$, and also the heavy fermion material Yb$_{1-x}$La$_{x}$Rh$_2$Si$_2$.  \cite{hayes2016scaling,giraldo2017scale,kumar2018high} The resistance is linear in both temperature and field. It is possible that generally linear-in-field and linear-in-temperature resistance go together in strongly correlated materials.  Studies of magnetoresistance in bad metals have not been common, and the data is not yet available to decide whether linear-in-field resistance is the norm in these materials.

In any case the discovery of linear-in-field resistance in the cuprates and pnictides unites the mystery of their linear resistance with the ongoing mystery of linear-in-field resistance in a wide variety of experimental realizations, many of which have Dirac cones.  \cite{zhang2012magneto,PhysRevLett.106.156808,PhysRevLett.108.266806,PhysRevB.92.081306,kisslinger2015linear,PhysRevB.92.041203,PhysRevB.89.125202,xu2015quasi,PhysRevB.93.104502,PhysRevB.84.100508,bhoi2011quantum}.  Several explanations of linear-in-field resistance have been given which focused on specific mechanisms which could be responsible for linear magnetoresistance in specific materials.  These include the cases of   ballistic conduction when the Fermi surface has a cusp, of a 3-D Dirac cone in the presence of a strong magnetic field, of a density gradient across the sample, and of classical transport with strong sample inhomogeneities. \cite{PhysRevLett.95.247201,PhysRevB.88.060412,PhysRevB.58.2788,PhysRevLett.117.256601,parish2003non}

In a recent work we gave a more generic explanation of linear resistance [linear both in field and in temperature]. \cite{sacksteder2018fermion} Our explanation is in terms of quantum interference; it is not specific to a particular microscopic mechanism such as details of the Fermi surface. It does, however, hint that the charge carriers responsible for linear resistance may be confined to move along locally one dimensional pathways. 

We argued that linear resistance is caused by the same combination of scattering and quantum interference which produces weak (anti)localization in ordinary materials.  Weak (anti)localization expresses the quantum interference which occurs when an electron after several scatterings returns to its starting position, forming a loop.  In this circumstance where an electron's sequence of scatterings forms a loop,  a special quantum interference will occur if a hole follows the same sequence of scatterings but in the opposite order.  The net effect is that disorder causes electrons and holes to form pairs called Cooperons.  It is  Cooperon loops [electrons following  loops in one direction and holes following in the opposite direction] that mediate weak (anti)localization and alter electronic conduction.

In ordinary 2-D and 3-D materials scattering causes quantum decoherence, so that when a Cooperon completes a loop, it is unable to repeat the same loop again.  This limit on Cooperon loops controls the  resistance signature of weak localization - for instance in  two dimensions the resistance varies logarithmically.    We argued that when linear resistance is measured, one can deduce from the linear resistance  that Cooperons are repeating their loops, and are not experiencing the quantum decoherence that prohibits loop repetitions in ordinary 2-D and 3-D materials.
 
The arguments of the above article were based entirely on analysis of the linear profile of the resistance, which does not in itself provide material for discerning which mechanism allows linear resistance materials to maintain quantum coherence even in the presence of scattering.  However one way of realizing robust quantum coherence, i.e. of allowing Cooperons to repeat their loops, is if conduction is confined to move along tracks that are locally one-dimensional.  There are several known ways to confine electrons  to quasi-1-D channels.  One such way is thin nanowires: if they are made long enough quantum interference takes over resulting in either Anderson localization or a perfectly conducting channel.  \cite{ishii1973localization,simon1989trace,RevModPhys.69.731,ando2002presence,takane2004quantum} Two other ways of confining electrons to quasi-1-D channels are the edges of topological insulators, and the snake states that are generated when graphene or TIs are subjected to various control mechanisms. \cite{PhysRevB.77.081403,PhysRevB.77.081404,PhysRevLett.107.046602}  In each of these 1-D realizations quantum interference is very robust, no matter how strong the scattering may be.  

In this respect we are reminded of very striking STM images of electron density on the surfaces of Ca$_{1.88}$Na$_{0.12}$CuO$_2$Cl$_2$, Bi$_2$Sr$_2$Dy$_{0.2}$Ca$_{0.8}$Cu$_2$O$_{8+\delta}$, \cite{kohsaka2007intrinsic} and Bi$_2$Sr$_2$(Ca,Dy)Cu$_2$O$_{8+x}$.  \cite{parker2010fluctuating} These experiments show electron density focused on line segments oriented along the bonds in the CuO$_2$ plane, which have two directions that are perpendicular to each other.  The line segments can be quite long, 15 bonds or more.  The STM image reveals that the plane is covered with these line segments, with apparently random length, angular orientation, and placement.  Perhaps charge conduction is guided along these quasi-1-D segments, producing the  quantum coherence necessary for linear resistance.

If the linear resistance seen in strongly correlated materials, including high $T_c$ materials above $T_c$, is caused by quasi 1-D conduction, then this is evidence that topology may be at work in these materials.  Topology is a very robust way of reducing dimensionality.

\section{Qualitative Similarities to Topological Insulators \label{TISimilarities}}
There are several qualitative similarities between high-$T_c$ superconductors and topological insulators.  

\textbf{Resilience Against Strong Scattering.} The first qualitative similarity is that in both classes of materials electronic conduction is resilient against very strong scattering.  Even very strong disorder is unable to stop topologically protected conduction on the surface/edge of a TI, as long as the bulk of the TI does not conduct. \cite{chen2012disorder,PhysRevB.85.195140,PhysRevApplied.3.064006} Unlike some TIs, high $T_c$ materials may not have strong static disorder.  However high $T_c$ materials, and strongly correlated materials, do have very strong scattering.  This is seen very clearly in ARPES studies of strongly correlated materials, which show that most of the spectral weight is not at the Fermi energy, but instead washed out into a smooth background.    The smooth background is evidence of very strong scattering.  It is remarkable, then, that the cuprates are very good superconductors even though they also show such very strong scattering.    The combination of very strong scattering with good conduction suggests that   high-$T_c$ superconductors may have something in common with TIs.

 \textbf{Linear Resistance.} A second qualitative similarity between high-$T_c$ materials and topological insulators is that both classes of materials are able to manifest linear resistance.  In fact,  of the materials where linear magnetoresistance has been reported, most either host Dirac fermions, or it has been suggested that they host Dirac fermions.  Some theoretical explanation for this comes from  Abrikosov's quantum theory of linear magnetoresistance, which builds on the foundational assumption that the linear resistance material  hosts 3-D Dirac electrons in its bulk.  Therefore when linear magnetoresistance is reported in strongly correlated materials, it is natural to look for links between those materials and topological physics.  
 
 This sort of reasoning, from linear resistance to  topology, has had some success in the case of certain pnictide superconductors where linear magnetoresistance is seen, and where both experimental evidence and theoretical work indicate the presence of Dirac cones and topological phases. \cite{PhysRevLett.105.037203,PhysRevLett.104.137001,bhoi2011quantum,PhysRevB.84.100508,PhysRevB.88.165402,pallecchi2013role,PhysRevB.90.024517,PhysRevB.93.104502,PhysRevB.93.104513,PhysRevB.96.220509,PhysRevLett.119.096401,PhysRevX.8.011014,zhang2019multiple,hao2019topological}  Much of this work belongs in the category of work that we here call band topology, since the theoretical work is largely based in band theory, and many of the experiments use techniques such as  quantum oscillations and ARPES which are essentially probes of band structure. \cite{PhysRevLett.104.137001,PhysRevB.93.104513,PhysRevB.96.220509,PhysRevX.8.011014,zhang2019multiple}.   We are suggesting here that the link from linear magnetoresistance to topology may have a broader relevance going beyond band-oriented analysis.  The pnictides are encouraging in this respect because they are strongly correlated, and their physics goes well beyond standard band structure, as witnessed by the ongoing discrepancies between ab initio predictions and experimental measurements of the bands. \cite{PhysRevB.94.201107,coldea2018key,rhodes2020revealing}
 
 % However the pnictides are strongly correlated materials, with physics that goes well be

\textbf{Weak Antilocalization and Fast Spin Relaxation.} The third and last qualitative similarity between high-$T_c$ materials and  topological insulators is that the linear resistance always goes up, not down, with magnetic field and with temperature.  This shared behavior supplies information about spin physics in both materials.   As outlined above,  in Ref. \onlinecite{sacksteder2018fermion} we have argued that the linear resistance is a manifestation of cooperon physics, the same physics which produces weak localization in ordinary materials.  Since the cooperon is a pairing of an electron with a hole, and since both the electron and the hole are spin $1/2$ particles with two states, the cooperon is a $2 \times 2$ matrix. The four numbers in the Cooperon matrix encode both  electronic charge [ a singlet ] and spin polarization [ a triplet ].  

When both the charge singlet and spin triplet components of the Cooperon have similar decay times, then the Cooperon causes the resistance to decrease with magnetic field  - this is called weak localization. \cite{hikami1980spin,PhysRevLett.48.1046}  In contrast, when the relaxation time of the Cooperon's spin triplet is short compared to the singlet's decay time, then the Cooperon causes the resistance to increase with magnetic field - this is called weak antilocalization.   When weak antilocalization is observed instead of weak localization, i.e. when the resistance increases not decreases with field, this is an unambiguous signal that the relaxation time of the Cooperon's spin triplet is quite short.  It must be so short  that the Cooperon's spin polarization relaxes to zero before a Cooperon is able to return to its starting point and form a loop.

The reason why in TIs the Cooperon's spin triplet has a very short relaxation time is that TIs have very strong spin-orbit couplings. \cite{PhysRevB.85.205303} In other words,  the spin state of an electron has a very strong impact on the direction of movement that is energetically preferred by the electron.  Whenever an electron scatters and therefore changes direction, it also changes its spin state.  In TIs the Cooperon's spin relaxation time is very close to the scattering time, i.e. much smaller than the time required for a Cooperon to return to its origin.   We turn briefly to the diffuson, the classical partner to the cooperon.  Both the diffuson and the cooperon have the same spin triplet relaxation time, so if the cooperon's spin relaxation time is short then so is the spin relaxation time of the diffuson.  In summary, when weak antilocalization is observed, the role of spin polarization in long distance conduction [either via the cooperon or via the diffuson] is negligible. \cite{PhysRevB.83.241304}

If the linear resistance seen in high-$T_c$ materials is really caused by the Cooperon, then it is a kind of weak antilocalization, not weak localization.  This in turn implies that the  relaxation time of the Cooperon's spin triplet is quite short - so short that spin polarization plays no part in long distance conduction.  Moreover some very fast-acting physics must be responsible for  Cooperon spin triplet relaxation in the high-$T_c$ materials.   These are very interesting qualitative similarities between the high-$T_c$ materials and TIs, concerning both spin physics and conduction.

\section{The Key Role in Linear Resistance of Loops and their Areas\label{LoopKeyRole}}

We continue our discussion of Cooperons, which we claim are responsible for the linear resistance  [linear both in field and in temperature] seen in high $T_c$ materials above the cuperconducting transition.  Cooperons  are electron-hole pairs whose pairing is induced by scattering centers.   Because of this pairing, they are able to conduct charge over distances and times that are much longer than the scattering length and time.  This is in contrast to single electrons or holes, whose phase is disrupted at each scattering event and therefore are unable to conduct at scales significantly exceeding the scattering scale.  The characteristic scales [both distance and time] of Cooperons are far longer than the scattering scale or the crystalline unit cell.    

Because of this difference in scales, the details of the individual scattering events suffered by a Cooperon, and the kinetics of a Cooperon's movement between scattering centers,  do not determine its contribution to weak (anti)localization.  These details belong to length and time scales that are much shorter than the Cooperon's scale, and therefore their only influence on the Cooperon are to determine the diffusion constant, relaxation times, and similar constants describing mixing of charge and spin.  Even the shape  traced in real space by a Cooperon loop does not matter, because a description of this shape would necessarily refer to individual scattering events.  These facts can be verified by computing the perturbative bubble diagram that describes the Cooperon.

There are only  three quantities that that affect the Cooperon's contribution to weak (anti)localization: the area of the loop which it traces, the statistics of which loop areas are more probable, and the relaxation times of its  charge singlet and the spin triplet components.  In materials showing weak antilocalization, such as linear resistance materials including high $T_c$ superconductors, the spin triplet relaxes very quickly leaving only the charge singlet, so there remain only two quantities which determine weak antilocalization: the loop area and the statistics of which loop areas are more probable.  This information can be encoded  in a probability density $P(A)$ giving the probability that a Cooperon loop has area $A$.  \cite{sacksteder2018fermion} 

In materials displaying weak antilocalization the area distribution $P(A)$ contains a complete description of  the WAL signature seen in  electrical conduction. Magnetic field $B$ couples to a Cooperon loop via a phase factor $\exp(\imath 2 \pi A B / \Phi_0)$, where $\Phi_0 = h/e $ is the magnetic flux quantum.  Therefore, up to a multiplicative constant, the magnetoconductance signal $G(B)$ caused by WAL   is equal to the sum of the phase factors of each Cooperon loop, i.e. $G(B) \propto \sum_{loops} \exp(\imath 2 \pi A_{loop} B / \Phi_0)$. This  can be rewritten in terms of the area distribution:
\begin{equation} \label{AreaDistribution}
G(B) \propto \int\,{dA}\, \exp(\imath 2 \pi A B / \Phi_0) \, P(A)
\end{equation}

This focus on  counting loops and their areas is a very peculiar aspect of weak antilocalization, and is very far from the feeling of most other condensed matter physics.   Loops are geometrical and topological objects,  and the area of a loop is also a geometrical quantity.  The shape of the loops  is not important, and neither are the details of the Hamiltonian that determines electronic motion.  This focus on geometry rather than ordinary physics is suggestive of topology, and in fact topological insulators are one the most significant realizations of weak antilocalization.

\section{Linear Relations to Two Dimensional Sheet Density\label{LinearRelations}} Of the two quantities determining weak antilocalization, i.e. Cooperon loop area and the number of  Cooperon loops with a particular area, only the former quantity area is dimensionful.   The inverse or conjugate of area is two dimensional sheet density.  Because of sheet density's status as  the conjugate of area, the special emphasis that weak antilocalization gives to area also applies to sheet density.

One example of the focus on sheet density is equation \ref{AreaDistribution}, which    is a Fourier transform that starts with a probability distribution  $P(A)$ that is a function of area, and ends with magnetoconductance $G(B)$ which is a function of magnetic field.  The reason why this equation is mathematically correct is that $B/\Phi_0$ has units of sheet density.  When  loops are the focus of attention, it is natural to expect not only their area but also sheet density to have a key role in determing experimental phenomena.

In this connection it is  striking that the cuprates display several linear relations that connect sheet density to other quantities.

\begin{enumerate}
\item \textbf{Resistance's Dependence on Temperature, Magnetic Field, and Hole Sheet Density.}
\begin{itemize}
\item Resistance is linearly related to magnetic field in La$_{2-x}$Sr$_{x}$CuO$_4$  (LSCO),  BaFe$_2$(As$_{1-x}$P$_x$)$_{2}$, and  Yb$_{1-x}$La$_{x}$Rh$_2$Si$_2$ at temperatures above $T_c$.  \cite{hayes2016scaling,giraldo2017scale,kumar2018high}  Equation \ref{AreaDistribution} shows that, in the context of weak antilocalization, magnetic field is equivalent to sheet density, so the linear relation between resistance and field can be understood as a linear relation between resistance and sheet density.
\item Resistance is linearly related to temperature in many bad metals, including the ones mentioned above where linearity in field has been reported.
\item Ref. \onlinecite{barivsic2013universal} found that the coefficient of linear-in-temperature resistance is inversely related to hole doping in four cuprates:  La$_{2-x}$Sr$_{x}$CuO$_4$ (LSCO), YBa$_2$Cu$_3$O$_{6+\delta}$ (YBCO), Tl$_2$Ba$_2$CuO$_{6+\delta}$ (Tl2201), and HgBa$_2$CuO$_{4+\delta}$ (Hg1201).  Since all four cuprates have the same copper oxide plane with the same cross-sectional area, the linear relation to hole doping is also a linear relation to the sheet density of holes.  Performing this conversion from hole doping $p$  to hole sheet density $\rho_{2D}$, Ref. \onlinecite{barivsic2013universal}'s result is $R_{xx}(T) \propto T / \rho_{2D}$. 
\item The coefficient of the resistance's dependence on temperature has been found to be proportional to the coefficient of the resistance's dependence on magnetic field.  Therefore Ref. \onlinecite{barivsic2013universal}'s result implies that $R_{xx}(B) \propto B / \rho_{2D}$, where $\rho_{2D}$ is hole sheet density.
\end{itemize}
\item \textbf{The Pseudogap Family of Temperatures that Decrease with Hole Sheet Density.} The characteristic temperatures  found well above $T_c$ in the cuprates, including the pseudogap temperature, are linearly related to the sheet density of holes.  As discussed earlier, the experimental data sets organize in four straight lines [four lines for LSCO, three or four for YBCO] that are highest at small doping and decrease with doping towards a meeting point near the the high-doping edge of the superconducting dome.  This suggests the presence of a sheet density that is high at low hole sheet density and decreases as hole sheet density increases.   
\item \textbf{Temperatures that Increase with Hole Sheet Density.} Ref. \onlinecite{sacksteder2020quantized}'s appendix A  reports six data sets with characteristic temperatures that increase as the hole sheet density increases.  Five of these, measured with ARPES, the thermoelectric power, and the magnetic susceptibility, seem to be direct manifestations of the density of mobile holes because they intercept $T=0$ near the low doping edge of the superconducting dome.  \cite{PhysRevB.79.140502,chatterjee2011electronic,kim2004two,ohsugi1991cu}  The sixth data set, a pairing temperature measured using magnetic hysteresis, extrapolates to  $T=0$ at a lower doping $p=0.023$, presumably because of a sensitivity to pinned holes. \cite{PhysRevLett.96.047002,PhysRevB.69.144508}
\item   \textbf{$T_c$'s Dependence on Superfluid Phase Stiffness.} In Ref. \cite{bozovic2016dependence}, Bozovic et al measured the 2-D superfluid phase stiffness   $\rho_s$ in LSCO thin films at many different dopings on the overdoped side of the phase diagram. Phase stiffness is  given in units of temperature, but this is artificial - the  definition of phase stiffness contains a factor of $k_B$.  Bozovic et al's phase stiffness experiments were probes not of temperature but instead of a 2-D [sheet] density, which was obtained by measuring the magnetic inductance between two coils.  The inductance is sensitive to the length scales of the coils themselves and of the superconducting order parameter in the superconducting film interposed between the coils.   

Bozovic et al reported that $T_c$ is linearly related to the $T=0$ value of $\rho_s$.  Bozovic et al's data also shows that in most of the superconducting sector of the phase diagram $\rho_s$ is linearly related to temperature $T$.
\end{enumerate}

In summary, in the bad metal regime above $T_c$ there are linear relations between hole sheet density, magnetic field, temperature as a thermodynamic quantity, and the several characteristic temperatures found above $T_c$ including the pseudogap temperature.  The linear resistance is controlled by ratios of these quantities.  Below $T_c$ there is a linear relation between superfluid phase stiffness $\rho_s$ and temperature, and $\rho_s$'s value at $T=0$ is linearly related to $T_c$.

These experimental results show some kind of equivalence between hole sheet density, superfluid phase stiffness which is a sheet density, magnetic field which is a sheet density after division by $\Phi_0$, temperature as a thermodynamic quantity,  and several characteristic temperatures including both the pseudogap temperature and also $T_c$.   The linearity of these relations suggests that the equivalence between these quantities may be true in a sense that is mathematically and physically precise. 

These results can also be read as saying that hole sheet density controls the phase diagram of the cuprates.   As we discussed earlier, sheet density is conjugate to area, and both are geometric quantities.  It is very striking to see a geometric quantity such as sheet density or area control the phase diagram.  This complete ascendance of geometry over all other kinds of physics, especially at temperatures above $T_c$, strongly suggests that topology is a key determinant of the cuprate phase diagram.

\section{Atomic Units and Numerology\label{AtomicUnits}}
The linear relations between sheet density and other quantities bear closer examination, because their units and the numerical values of their slopes indicate that in the cuprates the system of atomic units $m_e = e = \hbar =  1$, supplemented with $k_B = 1$, is somehow preferred over other systems of units.  

\subsection{Atomic Units are Preferred in the Cuprates}

There are two senses in which atomic units seem to be preferred by the cuprates.
Firstly, in atomic units energy, magnetic field, and temperature all have the same units as sheet density, so that linear relations between these quantities are natural in the sense of having slopes that are unitless.  
\begin{itemize}
\item The magnetic  flux quantum $\Phi_0 = 2 \pi \hbar / e$ is unitless with numerical value of $ 2 \pi$, giving magnetic field units of sheet density.   
\item In atomic units momentum $\vec{p}$ has units of inverse length, so energy $ E \propto |\vec{p}|^2/2m_e$ has units of sheet density. 
\item When $k_B = 1$ temperature is equivalent to energy, and has units of sheet density.
\end{itemize}  %, so that equation   \label{AreaDistribution} simplifies to $G(B) \propto \int\,{dA}\, \exp(\imath A B ) \, P(A) $ and magnetic field $B$ is clearly the conjugate of area.  
%Moreover the units of $k_B T$ are the same as $p^2 / 2m = ( \hbar^2 / 2 m x^2) $. The $1/ x^2$ is sheet density, so we obtain $T = \rho_{2D} /2 $ in atomic units. 
Linear relations between these quantities, such as the linear experimental relations which we have discussed above,  are more natural when they all have the same units.

The second sense in which  atomic units seem to be preferred by the cuprates is that in atomic units the linear relations have  slopes that are of order $O(1)$.  This contrasts with  other systems of units  where the slopes have numerical values that are different from $1$ by orders of magnitude.  In subsection \ref{SlopeNumericalValues} we will list the numerical values of these slopes. In some cases the slopes seem to have integer values, in atomic units.  

Moreover, in several cuprates and pnictides the value of $T_c$ at optimal doping is comparable to the sheet density of hole pairs, within a factor of ten - we give examples in subsection \ref{TcVsSheetDensity}.  

These numerical happenstances are especially remarkable because the definition of the system of atomic units  [as opposed to other units] was motivated by considerations that do not concern cuprate superconductivity.  Although excessive attention to numerical values can be simply numerology,  the discovery that so many slopes are of order one should stimulate thought about whether there may be some physical reason why atomic units might be preferred.  

\subsection{Significance of Atomic Units}
Atomic units do have a physical significance which distinguishes them from other systems of units.  Specifically, when the electron mass $m_e$ and the elementary charge $e$ are set to one, the only properties that electrons retain are their paths in space-time, and their spin.  (The charge operator simply counts electron paths, and has a subsidiary role.)

The space-time paths which we are discussing are those of Feynman's formulation of quantum mechanics as a sum over paths. \cite{feynman1965quantum}
  In this formulation an electron follows not just one path, but instead many paths which fully explore the sample in which the electron is moving.  All of these many paths are summed to determine the evolution of $|\psi \rangle $, the electronic state.   We are not discussing the semiclassical paths of an electron's average motion.  For instance, in a magnetic field an electron's average position executes well-defined circular loops around the axis of the magnetic field, and one can measure  this cyclotron motion. This is not the sort of path we are talking about.  Instead we are talking about the infinitely many quantum mechanical paths, tracing many complex trajectories and fully exploring the sample, which sum up to produce the average cyclotron motion.

Charge conservation tells us that no electron's space-time path begins or ends in isolation.  Instead when an electron is produced, a hole/positron is also produced, and likewise each electron is annihilated only when a hole is annihilated.  At each of the two endpoints of an electron's path one always finds endpoints of hole paths, and vice versa.  One can follow from an electron path to the path of one of its two adjoining holes, and from that hole's path to another electron path, etc.  If one keeps following from electron to hole, to electron, etc., eventually these paths will compose to form a loop in space-time. Loops are the only option because it is impossible for an electron or hole path to end in isolation.  In other words, electron paths should always be considered in conjunction with hole paths, taking account that they they always form loops in space-time.

Electrons and holes are geometric objects - they are simply space-time loops that carry spin. They are also topological objects, since loops can be knotted, braided, etc.  Atomic units accentuate this geometric and topological meaning.
% In a sense, atomic units are a statement that electron mass $m_e=1$ and charge  $-e=-1$ are quantized, which is true  

   In a material with fast spin-relaxation,  the electron is effectively spinless.  In this case space-time loops give a complete description of electrons and holes; they are purely geometric/topological objects, without any additional properties.  
   
   Setting $\hbar = 1$ also has a geometric significance - it focuses attention on the noncommutation between position/time on one hand and translations (of position or time) on the other hand.
   
     If in high $T_c$ materials atomic units are in some sense preferred over other systems of units, this is an indication that geometry, and topology, govern these materials.  
   
\subsection{Numerical Values of the Slopes\label{SlopeNumericalValues}}
In this subsection we report the numerical values of the slopes of the linear relations between temperature, magnetic field, hole sheet density, resistance, and superfluid phase stiffness. The point is that in atomic units these slopes are all of order $O(1)$, and some of the slopes may be integers.
\begin{enumerate}
\item \textbf{Resistance's Dependence on Temperature, Magnetic Field, and Hole Sheet Density.}
\begin{itemize}
\item  Ref. \onlinecite{barivsic2013universal} reports the coefficient of linear-in-temperature resistance in La$_{2-x}$Sr$_{x}$CuO$_4$ (LSCO), YBa$_2$Cu$_3$O$_{6+\delta}$ (YBCO), Tl$_2$Ba$_2$CuO$_{6+\delta}$ (Tl2201), and HgBa$_2$CuO$_{4+\delta}$ (Hg1201).   The authors report that the resistance is $R_{xx} / (k_B T) = \pi  \times 64 \, a_0^2 \times p^{-1}$, where $p$ is the hole doping. [This is after conversion of the data in Figure 5b to atomic units - in the article's units it is $R/(k_B T) = p^{-1} \times 2.6 \Omega /$ K ].  Using the dimensions of the CuO$_2$ plane \footnote{We use $13.14$ Angstroms and $a \approx 3.82$ Angstroms for the lattice parameters of LSCO.} to convert from hole doping to hole sheet density $\rho_{2D} $ obtains $R_{xx} =  3.8 \, k_B T\,/ \,\rho_{2D} $.  Defining the slope coefficient $\alpha = dR / d(k_BT) $, this is $R_{xx} =  \alpha \, k_B T $ and $\alpha = 3.8 / \rho_{2D}$. The numerical value of $3.8$ is remarkable for being of order $O(1)$, and for being the same as the integer $4$ within experimental errors.
\begin{itemize}
\item The factor of $3.8$ seen in $R_{xx} =  3.8 \, k_B T\,/ \,\rho_{2D} $ and $R_{xx} =  3.8 \, \mu_B  B\,/ \,\rho_{2D} $ is equal to the integer value $4$, within experimental errors.   Ref. \onlinecite{sacksteder2018fermion} accounts for the $4$ with a factor of $2$ from the $2$ in $k_B T =E = p^2/2m_e$.  Another  factor of $2$ comes from the difference between sheet density of carriers and sheet density of Cooper pairs.  However Ref. \onlinecite{sacksteder2018fermion} does not provide any argument excluding the possibility of additional multiplicative factors.
%In Ref. I gave a simple way of arguing for the observed  linear dependence on both temperature and magnetic field, putting the two quantities on the same footing.  In that paper's reasoning  $\gamma$  turns out to be the ratio of the effective mass of the charge carriers to the bare electron mass, i.e. $\gamma = m/m_e$.  Even if this does turn out to be true, it does not really resolve the problem of why $\gamma = 1$, since it opens the question of why the charge carriers should have $m=m_e$ in bad metals, i.e. what mechanism could be responsible for protecting the mass of the charge carriers.  
%\item Ref. \cite{barivsic2013universal}'s found that the coefficient of linear-in-temperature resistance is inversely related to hole doping in four cuprates:  La$_{2-x}$Sr$_{x}$CuO$_4$ (LSCO), YBa$_2$Cu$_3$O$_{6+\delta}$ (YBCO), Tl$_2$Ba$_2$CuO$_{6+\delta}$ (Tl2201), and HgBa$_2$CuO$_{4+\delta}$ (Hg1201).  Since all four cuprates have the same copper oxide plane with the same cross-sectional area, the linear relation to hole doping is also a linear relation to the sheet density of holes.  Performing this conversion from hole doping $p$  to hole sheet density $\rho_{2D}$, Ref. \cite{barivsic2013universal}'s result is $R_{xx}(T) \propto T / \rho_{2D}$. 
 %The arguments of Ref.  include a factor of $4$  which could take the place of the experimental $3.8$, but leave undecided an overall multiplicative constant [which would  be $3.8/4 \approx 1$ experimentally if the factor of $4$ is included].  
 \end{itemize}
\item Resistance also depends linearly on magnetic field $B$, i.e.  $R(B) = \beta \mu_B  B$.  Here $\mu_B$ is the Bohr magneton and equals $1/2$ in atomic units, and $\beta $ measures the slope of linear-in-magnetic-field resistance.  
\item The slope in temperature can be compared to the slope in magnetic field by calculating the ratio $\gamma = \alpha / \beta = (dR/d(k_BT)) / (dR/d(\mu_B B))$.  % In Ref.  I gave an argument that $\gamma$ is the ratio of the effective mass of the charge carriers to the bare electron mass, i.e. $\gamma = m/m_e$. 
\begin{itemize}
\item Recently an experiment on the high-$T_c$ pnictide superconductor BaFe$_2$(As$_{1-x}$P$_x$)$_{2}$ measured that this ratio $\gamma = \alpha / \beta = 1$ is  identical to one in atomic units, within the experimental error bar of $7\%$.  \cite{hayes2016scaling} This numerical value of unity is remarkable, and is specific to the choice of atomic units.
\item Further studies of the cuprate La$_{2-x}$Sr$_{x}$CuO$_4$ and of Yb$_{1-x}$La$_{x}$Rh$_2$Si$_2$ near optimal doping have found that $\gamma$ varied slowly with doping, with values between $0.7$ and $2.3$.    \cite{hayes2016scaling,giraldo2017scale}.  These values of $\gamma$ correspond to linear magnetoresistance coefficients between $1.5$ and $5.5$; $R_{xx} =  \left[1.5,5.5\right] \times \mu_B  B\,/ \,\rho_{2D} $.
\end{itemize}
\item If $\gamma = 1$, as found in BaFe$_2$(As$_{1-x}$P$_x$)$_{2}$,  then the magnetoresistance is $R_{xx} =  3.8 \, \mu_B  B\,/ \,\rho_{2D} $.  This number is remarkable for being of order $O(1)$.
%\item The coefficient of the resistance's dependence on temperature has been found to be proportional to the coefficient of the resistance's dependence on magnetic field.  Therefore Barisic's result implies that $R_{xx}(B) \propto B / \rho_{2D}$, where $\rho_{2D}$ is hole sheet density.
\end{itemize}
\item \textbf{The Pseudogap Family of Characteristic Temperatures that Decrease with Hole Sheet Density.}
Ref. \onlinecite{sacksteder2020quantized} gathers all experimental reports on characteristic temperatures above $T_c$ in LSCO and YBCO, and finds that they coincide on families of straight lines with quantized slopes.  Here we report the slopes of the lines, which are of order $O(1)$.  We also report the values of these characteristic temperatures at zero doping.  The ratio of the zero doping temperatures to the characteristic temperature $T = 6000$ K associated with the cuprate unit cell is of order $O(1)$.
\begin{itemize}
\item The pseudogap family in LSCO has slopes equal to $dT/d\rho_{2D}=-0.64 \, \times \, \left[ 1, \, 1/2, \, 1/3, \, 1/4 \right] $.
\item The pseudogap family in YBCO has slopes equal to $dT/d\rho_{2D}=-0.78 \, \times \, \left[ 1/2, \, 1/3, \, 1/4 \right] $.  There may  also be a higher pseudogap line with slope  $-0.78 \, \times 1$, but insufficient data is available for high temperatures in YBCO.
\item The pseudogap lines in both LSCO and YBCO extrapolate to zero hole density at temperature $T(\rho_{2D}=0)=1000$ K $\times  \left[ 1, \, 1/2, \, 1/3, \, 1/4 \right] $.  Converting to atomic units, this temperature is $T(\rho_{2D}=0)=1000$ K = $\frac{1}{6.0} \times \rho_{A}$, where  $\rho_A = A^{-1} = 6000$ K is the sheet density  associated with the unit cell of the cuprate copper oxide plane, which has area $A = 53 a_0^2$.   In the other words, the ratio of the pseudogap temperature at zero doping $T(\rho_{2D}=0)$ to the characteristic sheet density $\rho_{A}$ of the CuO$_2$ plane is a factor of $\frac{1}{6.0}$.
\end{itemize}
The numerical values of $-0.64, \, -0.78,  \, 1/6.0$ are specific to atomic units, and are remarkable for being of order $O(1)$.
% Intercept at 1000K for YBCO [or 1030 for LSCO].
%k_B= 8.617333262145 \times 10^-5 eV/K, 27.211386 Hartrees/eV
% 1000 K = 1000 K x 8.617333262145 \times 10^-5 eV/K x  (27.211386)^{-1} Hartrees/eV = 3.17 x 10^{-3} Hartrees = 1/ (316 a_0^2)
% Cuprate unit cell area: a=b=3.84 Angstroms / (0.529177 Angstroms/a_0); a \times b = 52.7 a_0^2 = 1/(1.90 x 10^-2 Hartrees) , 1/(a \times b) = 1.90 x 10^-2 Hartrees * 27.211386 eV / Hartree / (8.617333262145 \times 10^-5 eV/K)  = 
% =>  1000 K / cuprate unit cell area = 1000 K / (1/ab) = 3.17 x 10^{-3} Hartrees / (1.90 x 10^-2 Hartrees) = 3.17 / 19.0 = 0.167 = 6^{-1} => 1/(a \times b) = 6000 K.
% The pseudogap lines hits T=0 at p=0.26 for LSCO and  p=0.215 for YBCO, so divide by doping:
% LSCO : (0.26)^{-1} * (1000 K / cuprate unit cell area ) = 0.64
% YBCO: (0.215)^{-1} * (1000 K / cuprate unit cell area ) = 0.78
% If you use \hbar^2 / 2 m_e A = k_B T, then there is an additional factor of two.
%The characteristic temperatures  found well above $T_c$ in the cuprates, including the pseudogap temperature, are linearly related the sheet density of holes.  As discussed earlier, most of the experimental data sets organize in four straight lines [four lines for LSCO, three or four for YBCO] that are highest at small doping and decrease with doping towards a meeting point near the the high-doping edge of the superconducting dome.  This suggests the presence of a sheet density that is high at low hole sheet density and decreases as hole sheet density increases.   
\item \textbf{Temperatures that Increase with Hole Sheet Density.}  Ref. \onlinecite{sacksteder2020quantized} Appendix  reports six experimental  temperature data sets that rise linearly with hole sheet density.   Five of these extrapolate to $T=0$ near the underdoped limit of the superconducting dome, i.e. in the interval $p = \left[0.049,0.081\right]$, which suggests that these temperatures are direct manifestations of the sheet density of mobile holes.  A sixth data set extrapolates to  $T=0$ at a lower doping $p=0.023$, presumably because of a sensitivity to pinned holes.  \cite{PhysRevLett.96.047002} The slopes $dT/d\rho_{2D}$ of these data sets, in atomic units where they are unitless, are all of order $O(1)$:
\begin{itemize}
\item $dT/d\rho_{2D}=1.02 $ for a coherence temperature measured using angularly integrated ARPES. \cite{PhysRevB.79.140502} 
%21 - Hashimoto, 2009, AIARPES, probably crystals
%[6150. -498.]
% 6150 is the slope in Kelvin.
% 1/(ab) = 6000K
% 6150/6000 = 1.02
% \item Hashimoto, 2009. \cite{PhysRevB.79.140502} Angularly integrated ARPES. LSCO. The authors take the first derivative of the spectrum with respect to energy, and identify a peak in the first derivative.  The temperature reported here, which they call a coherence temperature, is a break in the temperature dependence of the peak position. We omit a data point at $p=0.15$ which fits well with the linear regression because they did not actually reach the reported temperature $T = 364$ K.    Roughly consistent with Ref. \cite{PhysRevLett.81.2124}. The slope is $6200$ K without the $p=0.15$ data point, or $4900$ K with it. % 21 - Hashimoto
\item $dT/d\rho_{2D}=0.27$ for a temperature where a Gaussian peak is extinguished in ARPES. \cite{chatterjee2011electronic} 
%79 - Chatterjee, 2011, ARPES on BSCCO, thin films and single crystals
%[1610.39944832 -116.37537484]
% 1610 is the slope in Kelvin.
% 1610/6000 = 0.27
% Chatterjee, 2011. \cite{chatterjee2011electronic} ARPES. Bi2212. Below this temperature the spectrum contains a sharp Gaussian peak, and above this temperature the peak is absent.  Unlike all other data discussed in this article, this data is obtained from Bi2212. % 79 - Chatterjee
\item $dT/d\rho_{2D}=0.35$ and $0.21$ for the beginning and end of a linear regime in the thermoelectric power. \cite{kim2004two}
%16 - Kim, 2004, break from high-T linear TEP at high doping, ceramic
% [2081.71428571 -156.48571429]
% 2080 is the slope in Kelvin.
% 2080/6000 = 0.35
% 1/(ab) = 6000K
% Kim, 2004.  \cite{kim2004two} Thermoelectric power.  LSCO. The temperature reported here marks a  break from the linear signal seen at high temperatures.  Here we plot only the dopings at $p=0.20$ and higher.   The slope is $2100$ K, about half of the slope in Ref. \cite{PhysRevLett.81.2124}. % 16 - Kim
%17 - Kim, 2004, break from low T linearity in TEP at high doping,  ceramic
%[1264.   -81.3]
% 1260/6000 = 0.21
% 1260 is the slope in Kelvin.
% 1/(ab) = 6000K
%Kim, 2004. \cite{kim2004two}  Thermoelectric power. LSCO. The temperature reported here marks a break from the  linear signal seen at low temperatures.  Here we plot only the dopings at $p=0.20$ and higher.  % 17 - Kim
\item $dT/d\rho_{2D}=0.13$ for a Weiss temperature measured using nuclear quadrupole resonance.  \cite{ohsugi1991cu} 
 %34 - Ohsugi, 1991, susceptibility - Weiss temperature, probably ceramic
% [751.43403442 -37.22562141]
% 750 is the slope in Kelvin.
% 1/(ab) = 6000K
% 750/6000 = 0.125
% Ohsugi, 1991.  \cite{ohsugi1991cu}   Nuclear quadrupole resonance. LSCO.  The temperature reported here is a Weiss temperature obtained by fitting the nuclear spin relaxation rate to a Curie-Weiss law. % 34 - Ohsugi
\item $dT/d\rho_{2D}=0.64$ for a pairing temperature detected using hysteresis in the low field magnetization. \cite{PhysRevLett.96.047002,PhysRevB.69.144508} 
%Panagopoulos - 2
%[3864.48598131  -90.28037383]
% 3860 is the slope in Kelvin.
% 1/(ab) = 6000K
% 3860/6000 = 0.64
% \item  Panagopoulos, 2006.   \cite{PhysRevLett.96.047002}  See also Ref. \cite{PhysRevB.69.144508} by the same group. LSCO. This temperature marks the  onset of hysteresis in the temperature dependence of the low field magnetization, which is probably a sign of pinned vortices and of pairing. The low-doping data from $p=0.03$ to $p  = 0.10$ nicely follows a straight line originating at $T=0, \, p=0.023$ and extending up to room temperature.  The small $p=0.023$ intercept may be caused by the observable's sensitivity to both pinned and mobile holes.  The slope is about $3900$ K, roughly comparable to the slopes of the LSCO ARPES data sets. \cite{PhysRevB.79.140502,PhysRevLett.81.2124}
\end{itemize}
% This is not a real temperature.
%32 - Ino, 1998, ARPES, crystal
%[4579.56900124 -333.16977483]
% Ino, 2009. \cite{PhysRevLett.81.2124} Angularly integrated ARPES. LSCO. The quantity reported here is a measure of  the width of the Fermi surface. Unlike all other data discussed in this article, this is an energy scale converted to temperature, not an experimental temperature.  The experiment was performed at $T =  18$ K.  The quantity reported here includes a factor of $1/\pi$ which might be able to be renormalized at will. We include this data set because the four data points between $p=0.074$ and $p=0.203$ rise linearly with doping and extrapolate to the underdoped edge of the superconducting dome.  We omit the $p=0.30$ point from the linear fit, and we omit the $p=0$ point altogether because Figure 2 in Ref. \cite{PhysRevLett.81.2124} shows $p=0$ data that seems to leave little ground for extracting a width. The slope is $4600$ K. % 32 - Ino
\item  \textbf{$T_c$'s Dependence on Superfluid Phase Stiffness.} Bozovic et al measured the phase stiffness $\rho_s$ as a function of temperature and doping.  \cite{bovzovic2016dependence} The phase stiffness is reported in units of Kelvin, but the actual experimental technique is a measurement of geometric quantities not temperature.  Therefore finding an equivalence between sheet density and temperature is a highly non-trivial result.
\begin{itemize}
\item  Figure 1 in Bozovic et al's Ref. \onlinecite{bovzovic2016dependence} demonstrates that $\rho_s$ is linear in $T$ through most of the phase diagram, except at high temperatures close to $T_c$.  Bozovic et al provided the data for Figure 1 as a supplement to the article.  Using this data, we calculate the slope $d\rho_s / dT$ by subtracting the value of $\rho_s$ at $T=10 \,K$ from its value at $T = 20\,K$ in all the odd-numbered samples [22 samples] with $T_c \geq 20\,K$. The maximum value of $\Delta \rho_s$ in this range is $23.53 \,K$ and the minimum value is $19.36\,K$. This gives the slope  $d\rho_s / dT \approx -2.15 \pm 0.2$, which is remarkable for being equal within experimental scatter to the integer value $d\rho_s / dT =-2$.   
\item $\rho_s(T)$ is linear in $T$ at most temperatures, but steepens near $T_c$.  This results in a linear relation between $T_c$ [at $\rho_s = 0$] and $\rho_s$'s value at $T=0$, which Bozovic et al report as $d\rho_{s0}/dT_c = (0.37 \pm 0.02)^{-1}  = 2.7 \pm 0.15$.   \cite{bovzovic2016dependence} 
\end{itemize}

\end{enumerate}

\subsection{Superconducting $T_c$ vs. Hole Sheet Density \label{TcVsSheetDensity}}
It is natural in atomic units to compare sheet density to temperature, and holes are key to cuprate superconductivity.  Therefore we examine whether the sheet density of hole pairs can be used to estimate $T_c$ at optimal doping.  For each of the following compounds we report the ratio of optimal $T_c$ to the sheet density of hole pairs $\rho_{holes}/2$.  We also report the same ratio, but with electron pairs rather than hole pairs, for several pnictides.  In each case the ratio is of order $O(1)$.
% Cuprate unit cell area: a=b=3.84 Angstroms / (0.529177 Angstroms/a_0); a \times b = 52.7 a_0^2 = 1/(1.90 x 10^-2 Hartrees) , 1/(a \times b) = 1.90 x 10^-2 Hartrees * 27.211386 eV / Hartree / (8.617333262145 \times 10^-5 eV/K)  = 
% =>  1000 K / cuprate unit cell area = 1000 K / (1/ab) = 3.17 x 10^{-3} Hartrees / (1.90 x 10^-2 Hartrees) = 3.17 / 19.0 = 0.167 = 6^{-1} => 1/(a \times b) = 6000 K.
% 6000K =  1/(3.17 x 10^{-6} x 52.7 a_0^2)
\begin{itemize} 
\item \textbf{YBCO and LSCO}:  The optimal doping is about $p=0.16$ and the in plane lattice spacing is about $3.84$ Angstroms; therefore the the sheet density of hole pairs is $\rho_{holes}/2=480$ K.  In comparison, in YBCO optimal $T_c$ is about $94.3$ K \cite{PhysRevB.73.180505}, and in LSCO it is $41.6$ K. \cite{bovzovic2016dependence} Taking the ratio of optimal $T_c$ to $\rho_{holes}/2$, one obtains  $0.20$ in YBCO and $0.09$ in LSCO. 
% 3.84 Angstroms is fairly standard - depending on the reference it might be 3.82 or similar changes.
% prb 73 180505 Liang Bonn and Hardy have Tcmax = 94.3 K and optimal T_c at p=0.16.
% Momono et al., Physica C 233, 395 (1994) gives LSCO T_c = 39K at p=0.152,0.161
%\item YBCO: At optimal doping $p=0.161$, $\rho_{pair} \approx 480$ K, vs. $T_c = 93.5 $ K.  
%\item LSCO: At optimal doping $p=0.161$, $\rho_{pair} \approx 480$ K, vs. $T_c = 38 $ K.  
% Cuprate unit cell area: a=b=3.84 Angstroms / (0.529177 Angstroms/a_0); a \times b = 52.7 a_0^2 = 1/(1.90 x 10^-2 Hartrees) , 1/(a \times b) = 1.90 x 10^-2 Hartrees * 27.211386 eV / Hartree / (8.617333262145 \times 10^-5 eV/K)  = 
% =>  1000 K / cuprate unit cell area = 1000 K / (1/ab) = 3.17 x 10^{-3} Hartrees / (1.90 x 10^-2 Hartrees) = 3.17 / 19.0 = 0.167 = 6^{-1} => 1/(a \times b) = 6000 K.
% 6000K =  1/(3.17 x 10^{-6} x 52.7 a_0^2)
% a=b=3.84 Angstroms => 6000K, 6000 * 0.16 * 0.5 = 480K
% 0.16 * 0.5 * 1000000/(3.17 * (3.84/0.529)^2) = 479K
\item \textbf{BSCCO}:  In Bi-2223  the optimal doping is near $p=0.17$ [the optimal doping for Bi-2212] and the in plane lattice spacing is near $5.41$ Angstroms; therefore the the sheet density of hole pairs is $\rho_{holes}/2=510$ K.  \cite{jean2000oxygen} In comparison, optimal $T_c$ is $108$ K at ambient pressure, and $T_c=136$ K at a pressure of $36$ GPa.   \cite{chen2010enhancement}Taking the ratio of optimal $T_c$ to $\rho_{holes}/2$, one obtains  $0.21$ at ambient pressure and $0.27$ under pressure. 
% This crystal unit cell has twice as many Cu's per unit cell so we need another factor of two.
% Bi-2212, 91K, a=5.41 Angstroms, p=0.17 from Physica C 339 (2000) , "Oxygen nonstoichiometry, point defects and critical temperature in superconducting oxide"
% the p=0.22 [but this was obtained in some heuristic way using a formula they like] and T=108K and a=5.44 Angstroms comes from "Correlation between oxygen excess density and critical transition temperature in superconducting Bi-2201, Bi-2212 and Bi-2223"
% the a=5.40 Angstroms was for Bi-2212, and I'm not sure which variant was used for the 36.4 GPa experiment.
% a= 5.40 Angstroms for Bi-2212, from "TEM analysis of planar defects induced by Ti doping in Bi-2212 single crystals"
% a=b=5.40 Angstroms => 3030K, 3030 * 0.22 *0.5 = 330 K 
% 0.17 * 0.5 * 2* 1000000/(3.17 * (5.41/0.529)^2) = 513 K
% Tc = 108K at ambient pressure and 136K \pm 10K at 36.4GPa, from "Enhancement of superconductivity by pressure-driven competition in electronic order"
% Bi2223, delta = 0.22, 110 K at ambient pressure, a=b 5.44 Angstroms , where is this from?
% Bi2223, delta = 0.22, 136K, 36.4 GPa, delta = 0.22, a=b 5.44 Angstroms , where is this from?
\item \textbf{HgBaCaCuO}: In Hg-1223 the optimal doping is near $p=0.13$ [the optimal doping for Hg-1201]  and the in plane lattice spacing is about $3.86$ Angstroms; therefore the the sheet density of hole pairs is $\rho_{holes}/2=385$ K. \cite{PhysRevB.59.7209,wu2013have} In comparison, optimal $T_c$ is $130$ K at ambient pressure, and $T_c=164$ K at a pressure of $45$ GPa.  \cite{schilling1993superconductivity,PhysRevB.50.4260}  Taking the ratio of optimal $T_c$ to $\rho_{holes}/2$, one obtains  $0.34$ at ambient pressure and $0.43$ under pressure. 
% for hg-1201 optimal doping is p=0.13, also a=3.89, PhysRevB.59.7209
% Hg-1223, a=3.93 Angstroms, from "Superconductivity above 130 K" in the Hg-Ba-Ca-Cu-O system.
% a=3.85-3.87 Angstroms, from "What have we learnt from the highest-Tc superconducting Hg-based cuprates?"
% PhysRevB.50.4260 reports 164K under pressure.
%0.13 * 0.5 * 1000000/(3.17 * (3.86/0.529)^2) = 385 K
%Hg-1245 & $110$ & $0.236$& $63$ & $400$ &  $0.31$    \\ %Hg-1245: x=0.236, 110K, gives x in the oxygen plane as 0.212 in one sample with Tc=108 and 0.236 in another with Tc=110, and the number of holes in the IP plane is 0.151 and ? in the other. , a,b,c not supplied by this ref.  a=b=3.86 A from wu2013have.
\item \textbf{FeSe on SrTiO$_3$}: We also discuss FeSe, which is a parent compound for the iron pnictide superconductors.  Bulk FeSe has $T_c = 9$ K, but when a monolayer of FeSe is supported by a SrTiO$_3$ substrate $T_c$ rises to $65$ K or higher.  The optimal doping is $p=0.12$ electrons per iron atom \footnote{Figure 2h of Ref. \onlinecite{ he2012phase} gives the optimal doping between $p=0.11$ and $p=0.12$ electrons per iron atom. } and the in plane lattice spacing is  $3.90$ Angstroms; therefore the the sheet density of electron pairs is $\rho_{electrons}/2=350$ K.  \cite{huang2017monolayer} In comparison, optimal $T_c$ is $65$ K or higher, although $T_c=109$ K has also been reported.   Taking the ratio of optimal $T_c$ to $\rho_{electrons}/2$, one obtains  $0.19$.
% a=3.90 comes from "Monolayer FeSe on SrTiO3" review by huang and hoffman, which mentions that by playing with the substrate one can get 75K.
% where did the p=0.12, and the T_c come from? These both came from the same article, "Phase diagram and high temperature superconductivity at 65 K in tuning carrier concentration of single-layer FeSe films", Fig 2h shows how Electrons/Fe changed with annealing cycles, reaching 0.11-0.12. 
% a=b=3.77 Angstroms => 6200K, 6200K * 0.12 *0.5 = 350 K
%0.12 * 0.5 * 1000000/(3.17 * (3.90/0.529)^2)
% annealed single-layer FeSe, 0.12 holes/Fe 65 K after many annealings, 
% a=3.77 and c=5.49 for bulk FeSe per the PNAS article " Superconductivity in the PbO-type structure 􏰀-FeSe"
\item \textbf{\textit{Re}FeAsO$_{1-x}$H$_x$(F$_x$) (\textit{Re}=La,Ce,Sm and Gd)}:  The members of this family of pnictides show similar doping vs. $T_c$ phase diagrams.  \cite{chen2014iron,takahashi2015superconductivity,hosono2018recent}  Here we concentrate on LaFeAsO$_{1-x}$H$_x$.  At a pressure of 6 GPa and an optimal doping of $x=0.18$ this compound reaches $T_c = 52$K.   $T_c$ is  weakly dependent on $x$ between $x=0.08$ and $x=0.40$. (In other members of the family very similar temperatures and dopings are found, but at ambient pressure not at $6$ GPa.) The in-plane lattice spacing is  $4.035$  Angstroms; therefore the sheet density of pairs is $\rho/2=490$ K.  Taking the ratio of optimal $T_c$ to $\rho/2$, one obtains  $0.11$ at optimal doping, or $0.24$ at the lower end  of the superconducting dome, or $0.05$ at the high doping end of the dome.
\end{itemize}

In summary, the sheet density of hole pairs overestimates the optimal $T_c$, but in the cuprates and pnictides the two numbers are often within an order of magnitude of each other.   In other words, $T_c/ (\rho_{holes}/2)$  is often between $0.1$ and $1$.   While this level of agreement is not satisfactory as a predictive tool, it is still an indicator that atomic units are preferred by the cuprates.   

The same point that optimal $T_c$ is of the same order as sheet density can be made using  the Roeser-Huber equation. \cite{roeser2008link,koblischka2020relation} This equation claims that there exists experimentally a simple relation between optimal $T_c$ and the spacing between charge carriers.  In atomic units the Roeser-Huber equation reads as $x^2 \times  T_c = 0.785 \times n^{2/3}$, where $x$ is a distance between charge carriers, so $1/x^2 = \rho_x$ is a sheet density.  $n=1,2,3$ is the number of  CuO$_2$ planes in the unit cell.   Reformulating the Roeser-Huber equation in terms of sheet density $\rho_x$ rather than carrier spacing $x$, it reads as  $T_c = 0.785 \times n^{2/3} \times \rho_x$.  In other words, it states that $T_c$ is linearly related to a sheet density $\rho_x$ by a coefficient of order $O(1)$.  The authors  who advanced the Roeser-Huber equation chose the value of $x$ for each material based on considerations about  the crystallography of the material. Agreement of a few percent  between the Roeser-Huber equation and experimental $T_c$'s was claimed for many cuprate and pnictide compounds.

Our point in mentioning the Roeser-Huber equation is: it could not work even within an order of magnitude, much less the claimed few percent, if $T_c / \rho_{2D}$ were not of order $O(1)$ in atomic units.  Although the case by case choice of the charge carrier spacing $x$ can be questioned, still the chosen values of $x$ were related to the lattice spacing, and therefore the tuning of $x$ cannot be responsible for the order of magnitude agreement of optimal $T_c$ and $ \rho_{2D}$.   The Roeser-Huber equation confirms that $T_c / \rho_{2D}$ really is of order $O(1)$ in atomic units.

\subsection{Significance of Integer Values.}
The slopes listed above include three numbers which are  very close to integer values:
\begin{enumerate}
\item In BaFe$_2$(As$_{1-x}$P$_x$)$_{2}$ the slope of the magnetic field dependence is the same as the slope of the temperature dependence; $\gamma = \alpha / \beta$ is  identical to one in atomic units, within the experimental error bar of $7\%$.  \cite{hayes2016scaling}
\item Bozovic et al's article showed that the superfluid phase stiffness $\rho_s$ is related to temperature by the integer $-2$ [within experimental scatter]; $d \rho_s /dT = -2$.   \cite{bovzovic2016dependence} 
\item Barisic et al report  that $R_{xx} = 3.8 k_B T / \rho_{2D}$.  $3.8$ is very close to the integer $4$, so perhaps the correct relation is  $R_{xx} = 4 \, k_B T / \rho_{2D}$.  \cite{barivsic2013universal} 
\end{enumerate}

When experimental data produces integer  values, this asks very strongly for our attention.  One must ask whether some physical mechanism guards the integer value from the many processes that could modify it.  One of the simplest and most convincing ways to make an integer value resilient against interactions and other processes is to use topology, for instance via a winding number.

\section{Meissner Physics \label{Meissner}}

It should be mentioned that even conventional superconductivity has  features that seem reminiscent of topology: a contrast between bulk physics and surface physics, and guaranteed conduction on the surface.  Where a TI  bulk band gap expels charge carriers from the bulk, superconductors expel magnetic field from the bulk.  In a TI charge conduction on the surface is guaranteed by topology and the bulk gap, while in a superconductor Meissner currents on the surface are required in order to expel magnetic field from the the bulk.

While a comparison  between topology and conventional superconductors is in some respects attractive, this comparison has two fundamental difficulties.  The first is that  superconductors can not be directly compared to TIs, since one  does not allow charge conduction in the bulk and the other does.  Perhaps the solution to this difficulty is think of Weyl materials instead of TIs.   The second and more profound difficulty is that this sort of reasoning can be used to suggest that ordinary conductors have topological features.  Ordinary conductors expel electric fields from their bulk, and use surface charge accumulation to achieve that expulsion.

\section{Postscript: How to do  topology in a strongly interacting material \label{Postscript}}
Band topology is not a terribly promising tool for understanding high $T_c$ materials, because these materials are so strongly interacting that the concepts of bands and Fermi surfaces are not reliable. Nonetheless the evidences reviewed in this article suggest that topology may be important.  Here we briefly outline a few possible directions for doing topology without band structure. This is not an exhaustive list.
\begin{itemize}
\item \textbf{Focus on Many-Particle Aspects.} One  approach is to attack strong interactions head-on: first write an interacting hamiltonian or a many-particle ground state, and then within this context work on developing a topological perspective.   One example of this approach is development of many-body topological invariants, where the idea is to develop mathematical machinery for analyzing a multiparticle ground state and determining whether it implements topological order.  For examples of this work see Refs. \onlinecite{senthil2015symmetry,RevModPhys.88.035005,PhysRevLett.118.216402}.

Another example is the Kitaev  model, whose solutions are spin liquids, and are intrinsically topological.  The Kitaev model can be solved exactly when it is on a honeycomb lattice. Depending on the lattice, the Kitaev model implements  nodal lines, Dirac cones, Majorana Fermi surfaces, or Weyl nodes.\cite{hermanns2018physics}

While this approach is perhaps the most obvious, it might not always be the most productive.  The challenges and technology of topology [either in the mathematical sense or the physics sense] are relatively well understood, in comparison to those of strong correlations and interactions. 
% The latter problem has not yet admitted either a clear formulation or a tractable set of technologies for its solution.  
Therefore attacking the combination of topology with strong interactions by concentrating on strong interactions first may deliver practical results more slowly than starting with the aspects of this combination that are relatively tractable.
\item \textbf{Focus on Broken Translational Invariance and the Absence of Band Structure.} Another approach  is to focus on how to do topology when translational invariance is broken by scattering and therefore there is no band structure.   The source of scattering can be static, i.e. disorder, or it can be dynamic, i.e. interactions.   If the characteristic time scale of interactions and their fluctuations is longer than that of electron motion, then the effects of the two types of scattering may be similar.

Topological properties can be extraordinarily resilient against scattering.  This can be observed by directly calculating topological invariants using formalisms which  admit formulations that are independent of band structure and can be calculated in disordered materials.  \cite{PhysRevB.85.075115,yi2013coupling,PhysRevLett.105.115501} 
% Coupling-matrix approach to the Chern number calculation in disordered systems, and PRL 105, 115501
% Computationally there  is a difficulty that computational scaling generally does not allow system sizes larger than $\approx 20-30$ lattice spacings, while in disordered systems getting crisp data on phase transitions etc. generally requires much larger system sizes.  However as  theoretical tools these techniques are important for understanding the combination of topology and scattering. 

 Resilience against scattering can also be observed by measuring conduction and the energy-resolved density of states, neither of which is tied to translational invariance.   The next three points discuss this in more detail:
\begin{itemize}
\item \textbf{Conduction.} Topology can guarantee conduction on the surface of a material even when scattering completely erases the band structure and band gap both on the surface and in the bulk; the only requirement is that the bulk must not conduct, which can be arranged by making it sufficiently disordered.   \cite{chen2012disorder,PhysRevB.85.195140,PhysRevB.85.035107}

Topology can also guarantee conduction in the bulk of a material, if the Fermi energy is tuned to a topological phase transition.  Because topological numbers cannot change without conduction through the bulk, at a topological phase transition the bulk is guaranteed to conduct.  \cite{chen2012disorder,PhysRevB.85.195140}
\item \textbf{Density of States as a Function of Energy.}  The characteristic signature of a Dirac cone  [constant density of states for 1-D edge states, linear in energy for 2-D surface states, quadratic  in energy for 3-D bulk states]  can be maintained by topology even when scattering is so strong that there is no dispersion relation between momentum and energy.  In this case the only effect of scattering on the density of states is to multiply by a constant, corresponding to multiplying the Fermi velocity by a constant. \cite{PhysRevB.88.045429,PhysRevApplied.3.064006} This result applies within a phase, while at boundaries between two phases non-trivial scaling relations may control the DOS. \cite{PhysRevLett.112.016402}
% H  = v_F k, E = v_F k, rho(E) = \Omega \int dk k^{D-1} delta(v_F k - E) = v_F^{-1} \Omega  (E/v_F)^{D-1} 
\item \textbf{Effective Models.} In topological insulators the effects of scattering on the  density of states and conduction can be summarized by renormalizing the parameters of a disorder-free Dirac hamiltonian.  In other words, even when translational invariance is completely absent and there is no dispersion correlating momentum with energy, topology's invisible hand can ensure that the material acts as if a scattering-free Dirac particle were present. \cite{PhysRevB.88.045429,PhysRevApplied.3.064006}

 Weyl materials take this a step further.  In 3-D Dirac materials disorder is an irrelevant operator, so its effect on the Weyl cone [except at the Weyl point itself] can be made arbitrarily small by  going to large enough distance scales and small enough momentum scales. \cite{PhysRevB.79.045321,PhysRevLett.107.196803}  For some observables such as conductance disorder may have no effect at all. \cite{PhysRevLett.110.236803}
 \end{itemize} 
\item \textbf{Renormalization Group Flow to Weyl Semimetals.} A material's properties at long length scales may be much different than its properties at short length scales.  For example, recent work has suggested that materials with spin-orbit coupling, magnetic impurities, and disorder may naturally flow  into a Weyl semimetal phase.  \cite{PhysRevB.98.205133}
Topological insulators with magnetic impurities can also flow into a Weyl semimetal phase as the system's length scale is increased. \cite{PhysRevB.91.115125}
This sort of scenario could  produce a Weyl phase and Fermi arcs in the cuprates even if at short distances there are no Dirac or Weyl fermions.

Moreover when stacks of 2-D TIs are coupled together into a 3-D material,  they can form Weyl semimetals or topological insulators. \cite{PhysRevB.94.235414, PhysRevB.92.235407}
If there were something topological at work in individual cuprate copper oxide layers, for instance  a Chern-Simons term or some kind of vortices, then possibly when the layers are stacked they could produce a Weyl phase and Fermi arcs. \cite{morinari2001d,morinari2003unified,PhysRevB.71.235102}
% d-Wave Superconductivity Induced by Chern-Simons Term in High-Tc Cuprates, Unified picture of Néel order destruction, d-wave superconductivity and the pseudogap phase for highT cuprates, also Phys. Rev. B 71, 235102
\item \textbf{Electron/hole loops.}
As briefly outlined earlier in this text, weak (anti)localization  is caused by Cooperons tracing loops.  We also briefly discussed the fact that electrons and holes form loops in space-time, which gives a perspective on electronic transport that is  founded on charge conservation and therefore is very solid.  These loops are an elementary topological object.  Studies of how the space-time loops of electrons and holes can be braided or knotted or woven together may  be able to offer new insight into strongly correlated electron transport.

We used an analysis of Cooperon loops and their areas to understand linear resistance. A key step in the analysis was that  weak antilocalization not weak localization is seen experimentally, which implies that spin relaxes very quickly, long before a loop is completed. This obtained a simplification: in the case of weak antilocalization, the loops are spinless.   Further work is needed on generalizing this loop  analysis to systems where spin polarization persists longer and its effects on transport can not be neglected.
\end{itemize}

 \begin{acknowledgments}  
We acknowledge the hospitality of J. Zaanen and the Instituut Lorentz in Leiden, A. Leggett and  the UIUC, and G. Kotliar and Brookhaven National Laboratory.  We are also thankful for helpful discussions with I. Bozovic, Jie Wu, A. Bollinger, I. Vishik, M. Hoesch, M. Watson, L. Rhodes, M. Norman, P. Johnson, S. Hayden, G. Parisi, P. Hasnip,  A. Petrovic, K.-S. Kim, A. Ho, J. Saunders, L. Levitin, J. Koelzer, T. Schapers, C. Beenakker, Alex Lau, Quansheng Wu, M. Sulangi, K. Schalm,  V. Cheianov, T. Ziman,    P. Coleman, and M. Ma.
 \end{acknowledgments}

\bibliography{Vincent}
%\subsection{Bibliography}

\end{document}